\documentclass{PoS}

\usepackage[utf8]{inputenc}
\usepackage{amsmath}
\usepackage{booktabs}
\usepackage{graphicx}
\usepackage{url}
\usepackage{gnuplot-lua-tikz}
\usepackage{macros}
\usepackage{macrosza}

\title{Non-perturbative improvement and renormalization of the axial current in $\nf = 3$ lattice QCD}

\ShortTitle{Improvement and renormalization of the axial current in $\nf = 3$ LQCD}

\author{\includegraphics[width=2.5cm]{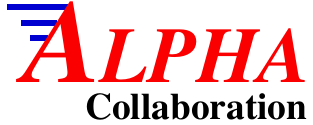}%
        \hfill%
        \raise22.35122pt\hbox{\footnotesize\it MS-TP-15-04}%
        \bigskip}

\author{John Bulava \\
        School of Mathematics, Trinity College, Dublin 2, Ireland \\
        E-mail: \email{jbulava@maths.tcd.ie}}

\author{Michele Della Morte \\
        CP$^3$-Origins \& Danish IAS, University of Southern Denmark, \\
        Campusvej 55, 5230 Odense M, Denmark \\
        E-mail: \email{dellamor@cp3-origins.net}}

\author{Jochen Heitger\footnotemark[1] \\
        University of Münster, Wilhelm-Klemm-Straße 9, 48149 Münster, Germany \\
        E-mail: \email{heitger@uni-muenster.de}}

\author{\speaker{Christian Wittemeier} \\
        University of Münster, Wilhelm-Klemm-Straße 9, 48149 Münster, Germany \\
        E-mail: \email{christian.wittemeier@uni-muenster.de}}

\abstract{We report on a non-perturbative computation of the
  renormalization factor $\za$ of the axial vector current in
  three-flavour $\Oa$~improved lattice QCD with Wilson quarks and
  tree-level Symanzik improved gauge action and also recall our recent
  determination of the improvement coefficient~$\ca$.  Our
  normalization and improvement conditions are formulated at constant
  physics in a Schrödinger functional setup.  The normalization
  condition exploits the full, massive axial Ward identity to reduce
  finite quark mass effects in the evaluation of $\za$ and correlators
  with boundary wave functions to suppress excited state contributions
  in the pseudoscalar channel.}

\FullConference{The 32nd International Symposium on Lattice Field Theory\\
                23-28 June, 2014\\
                Columbia University New York, NY}

\begin{document}

\section{Introduction}
\label{s:intro}

Wilson fermions~\cite{Wilson} are a popular way of discretizing
fermions in lattice QCD.  Their main drawback is that they break
chiral symmetry.  As a consequence, they are afflicted with
discretization errors linear in the lattice spacing~$a$.  These
$\Oa$~errors can be removed by adding the
Sheikholeslami-Wohlert term (clover term) to the Wilson
action~\cite{impr:SW} and further improvement terms to the matrix
elements of interest.  This method of $\Oa$~improvement has become
known as `Symanzik improvement program'~\cite{impr:Sym1,impr:Sym2}.
Moreover, currents which are conserved in a chiral theory and hence
need not be renormalized require a finite renormalization with Wilson
fermions.  One of these currents is the isovector axial current which
we are concerned with in this work.

The isovector axial current is a quark bilinear and in its bare form
it can be written as
\begin{align}
  A_\mu^a(x) & = \psibar(x) \, \gamma_\mu \gamma_5 \, \frac{\tau^a}{2} \, \psi(x),
\end{align}
where $\tau^a$ is a matrix acting in flavor space.  Symanzik's
effective theory predicts that the axial current will mix with the
derivative of the pseudoscalar density~$P^a(x)$ at $\Oa$, when the
lattice discretization breaks chiral symmetry.  This can be
compensated for by adding a corresponding improvement term to the bare
current.  Its coefficient is denoted by $\ca$ and is at the heart of
improving the axial current.

Furthermore, the axial current is renormalized by multiplying it by
the renormalization factor~$\za$ and another mass-dependent term.
Thus, the fully renormalized and improved axial current on the lattice
is
\begin{align}
  (\ar)_\mu^a(x) & = \za \left( 1 + \ba \, a \mq \right) \left[ A_\mu^a(x) + a \, \ca \, \drv\mu P^a(x) \right] && \text{with} & P^a(x) = \psibar(x) \, \gamma_5 \, \frac{\tau^a}{2} \, \psi(x).
\end{align}

The axial current plays a fundamental rôle in many applications,
notably the computation of quark masses and meson decay constants in
the pseudoscalar sector.  These are not only of phenomenological
interest, but they provide a precise way of setting a physical scale
in lattice simulations, too.  One observable that can be used for this
purpose is the kaon decay constant~$\fk$~\cite{lambda:nf2}.  In these
contexts, it is crucial to employ the improved and renormalized
current, since, typically, improvement and renormalization each
contribute about 10-20\% to the final
result~\cite{impr:ca_nf2,Kaneko:2007wh,impr:za_nf2}.  Both can be
computed in perturbation theory.  However, in previous works it was
found that the non-perturbative results deviate strongly from the
1-loop estimates.  The deviations can be several times the 1-loop
contribution itself.  Therefore, it is desirable to determine $\ca$
and $\za$ in a non-perturbative way.

The methods we use to determine $\ca$ and $\za$ have been introduced
in previous papers, which applied them to the
quenched~\cite{impr:pap3,impr:pap4} and two-flavor QCD
cases~\cite{impr:ca_nf2,impr:za_nf2}.  Here, we look at an action with
$\Oa$ improved $\nf = 3$ mass-degenerate dynamical Wilson
fermions~\cite{Bulava:2013cta} and the tree-level Symanzik-improved
gauge action (TLI gauge action)~\cite{Luscher:1984xn}.  The main part
of this text is about the renormalization of the axial current.  The
renormalization condition is summarized in \sect{s:condition}, the
ensembles of gauge field configurations that we used are described in
\sect{s:simulations}, and our preliminary results, in particular for
the interpolating function for $\za$ (valid for lattice spacings
$\lesssim 0.09\,\fm$), can be found in \sect{s:results}.  The
determination of the improvement coefficient~$\ca$ was recently
finished and will be published soon~\cite{impr:ca_nf3}.  We will only
reproduce the main result in \sect{s:results}.  A preliminary report
can also be found in \cite{lat13:ca_nf3}.

\section{Renormalization Condition}
\label{s:condition}

Renormalization conditions for the axial current are based on the idea
that they can restore chiral Ward identities, which are broken by the
Wilson term, up to $\Oasq$.  This is done by choosing one particular
Ward identity and adjusting $\za$ so that it holds exactly.  The
condition that we choose has been introduced for simulations with two
dynamical fermions~\cite{impr:za_nf2}, but a similar condition has
already been used in the quenched case~\cite{impr:pap4}.  We give a
shortened description of its derivation below.  More details can be
found in the original papers.

The Ward identity our renormalization condition is based on is
similar to the PCAC relation, i.e., it is derived from a chiral
rotation of the quark fields, but, in addition, the axial
current~$A_\nu^b(y)$ is inserted as an internal operator.  The
resultant identity is
\begin{align}
  \label{e:ward1}
  \int_{\partial R} \dd{}{\sigma_\mu(x)} \mvl{A_\mu^a(x) A_\nu^b(y) \opext} - 2 m \int_R \dd{4}{x} \mvl{P^a(x) A_\nu^b(y) \opext} & = i \epsilon^{abc} \mvl{V_\nu^c(y) \opext},
\end{align}
where $R$ is an arbitrary region containing $y$, $\opext$ is
an operator built from fields outside $R$, and $V_\nu^c$ is the
isovector vector current.  As region $R$ we choose the spacetime
volume between two space-like hyperplanes.  Furthermore, \eq{e:ward1}
can be modified by using the PCAC relation to shift the integration
domains.  Setting $\nu = 0$ and contracting the flavor indices~$a$ and
$b$ with $\epsilon^{abc}$, one arrives at
\begin{multline}
  \label{e:ward2}
  \int \dd{3}{\vecx} \dd{3}{\vecy} \epsilon^{abc} \mvl{A_0^a(x) \, A_0^b(y) \, \opext} \\
  {} - 2 m \int \dd{3}{\vecx} \dd{3}{\vecy} \int_{y_0}^{x_0} \dd{}{x_0} \epsilon^{abc} \mvl{P^a(x) \, A_0^b(y) \, \opext}
  \\ = i \int \dd{3}{\vecy} \mvl{V_0^c(y) \, \opext}
\end{multline}
with $x_0 > y_0$ defining the hyperplanes.

We evaluate this identity on a lattice with Schrödinger functional
boundary conditions (periodic in space, Dirichlet in
time)~\cite{SF:LNWW,SF:stefan1} with vanishing background field.  The
source operator $\opext$ is built from the quark fields
$\zeta$ and $\zetaprime$ at the boundaries $x_0 = 0$ and $x_0 = T$:
\begin{align}
  \label{e:opext}
  \opext & = -\frac{1}{6 L^6} \epsilon^{cde} \opprime{}^d \op^e
\end{align}
with
\begin{align}
  \label{e:sourceops}
  \op^e            & = a^6 \sum_{\vecu, \vecv} \zetabar(\vecu) \, \gamma_5 \, \frac{\tau^e}{2} \, \omega(\vecu-\vecv) \, \zeta(\vecv) && \text{and} &
  \opprime{}^d     & = a^6 \sum_{\vecu, \vecv} \zetabarprime(\vecu) \, \gamma_5 \, \frac{\tau^e}{2} \, \omega(\vecu-\vecv) \, \zetaprime(\vecv),
\end{align}
where the wavefunction~$\omega$ is understood to be an approximation
of the pseudoscalar ground state.  Its construction is detailed in
\cite{impr:ca_nf3,lat13:ca_nf3}.  We summarize the results in
\sect{s:results}.  The free index~$c$ that appears in \eq{e:opext} is
contracted with the free index from \eq{e:ward2}.  In this case, the
term on the right-hand side involving the vector current can be
simplified to the boundary-to-boundary correlator
\begin{align}
  F_1 & = -\frac{1}{3 L^6} \mvl{\opprime{}^a \op^a}
\end{align}
up to $\Oasq$, as was shown in~\cite{impr:pap4,impr:za_nf2} by using
isospin symmetry.

In the remaining two terms on the left-hand side, the continuum
currents are replaced by their improved and renormalized counterparts
from the lattice.  The result can be put in this form,
\begin{align}
  \label{e:ward3}
  \za^2 \left( 1 + \ba \, a \mq \right)^2 \left[ \faa^\I(x_0, y_0) - 2 m \cdot \fpa^\I(x_0, y_0) \right] & = F_1,
\end{align}
with the improved correlation functions
\begin{align}
  \label{e:faai}
  \faa^\I(x_0, y_0) & = \faa(x_0, y_0) + a \ca \left[ \tilde\partial_{x_0} \fpa(x_0, y_0) + \tilde\partial_{y_0} \fap(x_0, y_0) \right] \nonumber \\
                    & \qquad + a^2 \ca^2 \, \tilde\partial_{x_0} \tilde\partial_{y_0} \fpp(x_0, y_0), \\
  \intertext{and}
  \label{e:fpaitilde}
  \fpatilde^\I(x_0, y_0) & = a \sum_{x'_0=y_0}^{x_0} w(x'_0) \left[ \fpa(x_0, y_0) + a \ca \, \partial_{y_0} \fpp(x_0, y_0) \right],
\end{align}
where $\tilde\partial$ denotes the central difference operator and
$F_{XY}(x_0, y_0)$ with $X, Y \in \{A_0, P\}$ stands for
\begin{align}
  \label{e:fxy}
  F_{XY}(x_0, y_0) & = -\frac{a^6}{6 L^6} \sum_{\vecx, \vecy} \epsilon^{abc} \epsilon^{cde} \mvl{\opprime{}^d X^a(x) Y^b(y) \op^e}
\end{align}
and
\begin{align}
  w(x'_0) & = \begin{cases}
                1/2 & \text{if $x'_0 = y_0$ or $x'_0 = x_0$} \\
                1   & \text{if $y_0 < x'_0 < x_0$}
              \end{cases}
\end{align}
implements the trapezoidal rule.  Note that in \eq{e:ward3} the
renormalization factors, which would arise from the boundary
fields~$\zeta$ and $\zetaprime$, cancel on both sides and that the
product $m P^a$ can be renormalized with the same factor~$\za$ as the
axial current due to the PCAC relation.  The mass-dependent term
proportional to $\ba$ will be dropped from here on, since we will
impose the renormalization condition at vanishing mass.  Of course, we
can not tune the parameter to get exactly zero mass, but this only
amounts to an $\Or(am)$ effect.  Thus, our final renormalization
condition is
\begin{align}
  \label{e:renormalizationcondition}
  \za & = \lim_{m \to 0} \left[ \frac{F_1}{\faa^\I(x_0, y_0) - 2 m \cdot \fpa^\I(x_0, y_0)} \right]^{\frac{1}{2}}.
\end{align}
In order to maximize the distance between the insertion points, we
choose $x_0 = \frac{2}{3} T$ and $y_0 = \frac{1}{3} T$.

\begin{figure}
  \centering
  \includegraphics{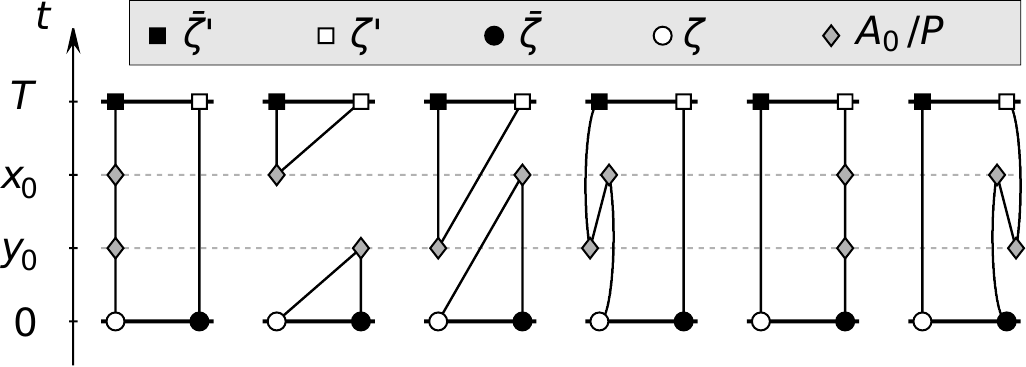}
  \caption{The six non-vanishing Wick contractions of correlation
    functions~$F_{XY}$ with source $\op^\mathrm{ext}$ and two bulk
    insertions $X$ and $Y$, see \protect\eq{e:fxy}, taken
    from~\cite{impr:za_nf2}.}
  \label{f:contractions}
\end{figure}

Except for $F_1$, \eq{e:renormalizationcondition} is built from
correlators of the form given in \eq{e:fxy}.  When one evaluates these
by performing the Wick contractions, one finds that only six
contractions contribute, which are illustrated in
\fig{f:contractions}.  Two of them are disconnected.  As argued in
\cite{impr:za_nf2}, they only give rise to $\Oasq$ contributions and
cancel in the continuum limit.  By omitting them and taking only the
connected contractions, one obtains an alternative renormalization
condition.  We will denote the corresponding renormalization factor by
$\zacon$.

\section{Simulations}
\label{s:simulations}

To determine the renormalization factor~$\za$ of the axial current, we
re-use the ensembles that were generated to obtain the improvement
coefficient~$\ca$.\footnote{The ensembles were generated using the
  openQCD code~\cite{algo:openQCD}, see also
  \url{http://luscher.web.cern.ch/luscher/openQCD/}.}  These had
already been designed to fit this purpose, in particular the ratio of
the spatial and temporal extents was approximately $T/L \approx 3/2$
with $L \approx 1.2\,\fm$.  From \cite{DellaMorte:2008xb}, we expect
this to be a good trade-off between a large infrared cutoff and small
$\Oasq$ effects.  An overview of the simulation parameters is given in
\tab{t:sim}.

\begin{table}
\centering
\setlength{\tabcolsep}{3pt}
\begin{tabular}{cccccc}
\toprule
$L^3\times T / a^4$ & $\beta$ & $\kappa$ & \#\,REP & \#\,MDU & ID   \\
\midrule
$12^3\times 17$     & 3.3     & 0.13652  & 10      & 10240   & A1k1 \\
		    &         & 0.13660  & 10      & 13048   & A1k2 \\[1ex]
$16^3\times 23$     & 3.512   & 0.13700  & 2       & 20480   & B1k1 \\
                    &         & 0.13703  & 1       & 8192    & B1k2 \\
                    &         & 0.13710  & 3       & 24560   & B1k3 \\[1ex]
$20^3\times 29$     & 3.676   & 0.13700  & 4       & 15232   & C1k2 \\
                    &         & 0.13719  & 4       & 15472   & C1k3 \\[1ex]
$24^3\times 35$     & 3.810   & 0.13712  & 7       & 15448   & D1k1 \\
\bottomrule
\end{tabular}
\caption{%
Overview of simulation parameters, number of replica and total number of 
molecular dynamics units of gauge configuration ensembles labeled by
`ID'.
}\label{t:sim}
\end{table}

The coupling~$\beta$ was chosen such that the physical lattice
size~$L$ stays roughly constant (line of constant physics).  In this
way, $\Oasq$ ambiguities in $\za$ are guaranteed to smoothly vanish in
the continuum limit.  This was done using the perturbative relation
between the lattice spacing~$a$ and the bare coupling.  However, only
the first two, universal coefficients~$b_0$ and $b_1$ of the
beta-function could be taken into account, because higher-order terms
are not known for the TLI gauge action.  To test for deviations from
the line of constant physics, the gradient (or Wilson) flow
coupling~$\gbarGF$ was computed~\cite{flow:FR}.  It is a renormalized
coupling that depends on $L$ as a scale, i.e., it will be constant if
$L$ is constant.

The parameter~$\kappa$ was tuned towards a vanishing PCAC mass.
In~\cite{Bulava:2013cta}, an upper bound of $|a \mpcac| <
0.015$ was employed.  We expect that $\za$ is more sensitive to the
mass.  Since at most values of $\beta$ we have several ensembles with
different $\kappa$ values, we can check its influence explicitly by
comparing the results.

In order to monitor the autocorrelation of the generated gauge
configurations, several observables were computed alongside which are
defined in terms of the gauge field smoothed by means of the gradient
flow~\cite{flow:ML,flow:LW}.  Mostly, they showed autocorrelation
times that did not exceed $250\,\mathrm{MDU}$, only the topological
charge became frozen at the largest $\beta = 3.810$.  For $\ca$ and
$\za$, this should amount to a cutoff effect, but we have estimated it
explicitly for $\ca$ and have indeed found no significant
deviations~\cite{impr:ca_nf3}.  Further details, in particular about
the algorithmic details can be also found in that reference.

\section{Results}
\label{s:results}

We have measured the correlators that are necessary to compute the
renormalization factor and the PCAC mass on every second trajectory
(in A1k2, A2k1) or every fourth trajectory (in the remaining
ensembles).  Via \eq{e:renormalizationcondition}, we compute $\za$ as
well as the alternative $\zacon$, where only the connected
Wick contractions are included.

As already anticipated in \eq{e:sourceops}, our Schrödinger functional
correlators involve boundary operators with a particular choice of
wavefunction, which is constructed in such a way that it suppresses
the contribution of the first excited state in the pseudoscalar
channel.  This optimal wavefunction was determined in the context of
our non-perturbative calculation of the improvement coefficient $\ca$
in~\cite{impr:ca_nf3} (see also \cite{lat13:ca_nf3} for a preliminary
report), which employs the same gauge field ensembles at constant
physics and the same kinematical setup as used here.  It relies on
demanding the quark mass extracted from the PCAC Ward identity to stay
unchanged when the external states are varied, where in practice these
external states are modeled as superpositions of spatial trial
(hydrogen-like) wavefunctions designed to approximately maximize the
overlap with the ground and first excited state, respectively.  For
the purpose of $\za$, however, only the wavefunction for the
approximate pseudoscalar ground state is required, which we therefore
choose as the one already obtained in~\cite{impr:ca_nf3}, i.e.,
\begin{align}
  \omega_{\pi^{(0)}} & = \sum_{i=1}^3 \eta^{(0)}_i \omega_i, &
  \eta^{(0)} & = (0.5317, 0.5977, 0.6000),
\end{align}
in terms of the basis of suitable trial wavefunctions $\omega_i$
mentioned above.  With this wave function at hand, the
\eq{e:renormalizationcondition} for $\za$ is evaluated upon prior
projection of all entering correlation functions to this approximate
ground state.

\begin{figure}
  \centering
  \input{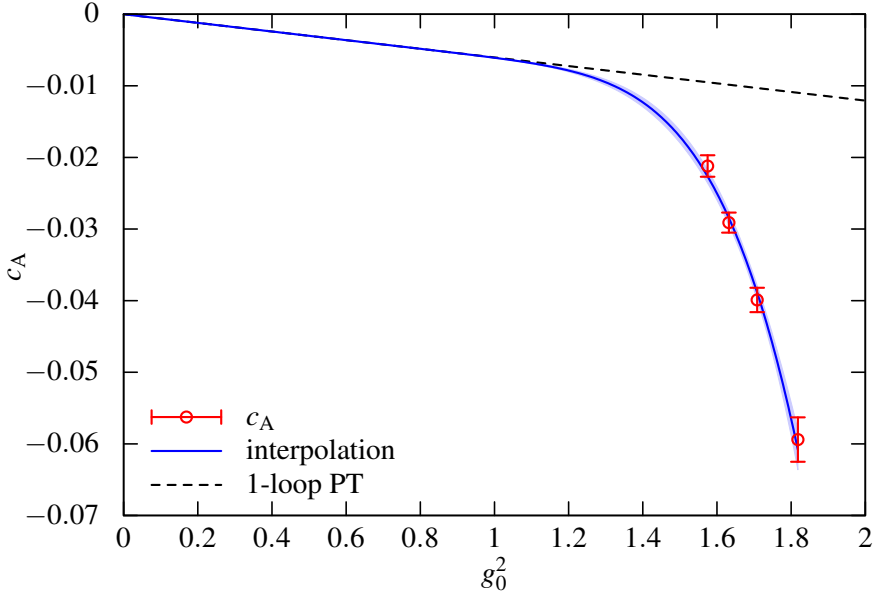}
  \caption{Final results for $\ca$ together with
    interpolation~\cite{impr:ca_nf3}.  The dotted line indicates the
    perturbative 1-loop asymptotics~\cite{impr:csw_iwa_pert}.}
  \label{f:ca}
\end{figure}

Moreover, as it is evident from \eqs{e:faai} and~(\ref{e:fpaitilde}),
a genuine non-perturbative determination of $\za$ also requires the
knowledge of non-perturbative values for $\ca$ for our simulation
parameters.  We thus rely on the result of our aforementioned recent
non-perturbative computation of $\ca$ in three-flavor lattice QCD with
tree-level improved gauge action~\cite{impr:ca_nf3}, which we
reproduce here for convenience:
\begin{align}
  \ca(\gosq) & = -0.006033 \, \gosq \times \left[ 1 + \exp\left(9.2056 - 13.9847 \cdot g_0^{-2} \right)\right];
\end{align}
this formula is valid for bare couplings below $\gosq \approx 1.8$
and with statistical errors between $\approx 4\%$ near the largest and
$\approx 8\%$ near the smallest bare couplings simulated.
For further details about this determination of $\ca$ we refer to
refs.~\cite{impr:ca_nf3,lat13:ca_nf3}.

Our preliminary results for $\za$ obtained for the ensembles
considered so far are collected in \tab{t:za}.  In our tentative error
analysis, we estimate the errors on $\za$ for each replicum via a full
autocorrelation analysis as described in \cite{Wolff:2003sm} and
compute a weighted average over all replica within an ensemble.

\Tab{t:za} also includes the unrenormalized PCAC mass~$a
\mpcac$, which is computed using the perturbative value of
$\ca$ and the wavefunction~$\omega_{\pi^{(0)}}$, the gradient-flow
coupling $\gbsqGF$ and the results for $\zacon$ obtained via
the alternative definition of the renormalization factor, which
includes only connected contractions.

As can be seen from the table, $\gbsqGF$ is approximately constant,
only the deviation on the ensemble D1k1 at the largest $\beta$ is more
pronounced.  We do not expect that this deviation from constant
physics has a significant effect on our result, but we plan to check
it explicitly using the ensemble B2k1 from \cite{impr:ca_nf3}, whose
parameters are identical to the ones of B1k1 except for $\beta$.  Some
of the ensembles in the B and C groups show a significant though not
yet severe mass dependence.  We consider taking a closer look at this
issue and adding new ensembles with different $\kappa$ values, too.

For $L/a = 12$ ($\beta = 3.3$), $\zacon$ differs
significantly from the standard value~$\za$, which seems to signal
significant $\Oasq$ uncertainties in $\za$ at this lattice spacing ($a
\approx 0.09\,\fm$).  Similarly large cutoff effects at this $\beta$
were also observed for this action in~\cite{Bruno:2014jqa}.  However,
at smaller lattice spacings the results for both definitions of $\za$
are in good agreement within their errors.  We plan to examine the
impact of cutoff effects on $\za$ more closely by adding an ensemble
at $L/a = 14$ ($\beta = 3.414$).

\begin{table}
  \centering
  \begin{tabular}{llllll}
    \toprule
    ID   & $a \mpcac$    & $\gbsqGF$ & $\zacon$   & $\za$      &        \\
    \midrule
    A1k1 & $-0.0010(7) $ & 18.12(21) & 0.8162(78) & 0.6553(90) & $\ast$ \\
    A1k2 & $-0.0086(6) $ & 16.95(13) & 0.8290(92) & 0.6489(72) &        \\[1ex]
    B1k1 & $+0.0063(2) $ & 16.49(13) & 0.7757(23) & 0.7666(47) &        \\
    B1k2 & $+0.0056(3) $ & 16.85(20) & 0.7758(45) & 0.7677(71) &        \\
    B1k3 & $+0.0022(2) $ & 16.11(14) & 0.7804(30) & 0.7516(36) & $\ast$ \\[1ex]
    C1k2 & $+0.0066(2) $ & 15.53(14) & 0.7889(16) & 0.7888(48) &        \\
    C1k3 & $-0.0005(1) $ & 14.64(13) & 0.7822(27) & 0.7785(33) & $\ast$ \\[1ex]
    D1k1 & $-0.00269(8)$ & 13.90(11) & 0.7969(16) & 0.7904(16) & $\ast$ \\
    \bottomrule
  \end{tabular}
  \caption{Summary of results: the unrenormalized PCAC quark mass, the
    gradient-flow coupling, the results for $\za$ using the
    alternative definition ($\zacon$) and the standard
    definition including disconnected contractions.  Ensembles marked
    by `$\ast$' are used in the fit procedure.}
  \label{t:za}
\end{table}

\begin{figure}
  \centering
  \begin{tikzpicture}[gnuplot]
\path (0.000,0.000) rectangle (12.000,8.000);
\gpcolor{color=gp lt color border}
\gpsetlinetype{gp lt border}
\gpsetlinewidth{1.50}
\draw[gp path] (1.504,0.985)--(1.684,0.985);
\draw[gp path] (11.447,0.985)--(11.267,0.985);
\node[gp node right] at (1.320,0.985) {$0.64$};
\draw[gp path] (1.504,1.650)--(1.684,1.650);
\draw[gp path] (11.447,1.650)--(11.267,1.650);
\node[gp node right] at (1.320,1.650) {$0.66$};
\draw[gp path] (1.504,2.314)--(1.684,2.314);
\draw[gp path] (11.447,2.314)--(11.267,2.314);
\node[gp node right] at (1.320,2.314) {$0.68$};
\draw[gp path] (1.504,2.979)--(1.684,2.979);
\draw[gp path] (11.447,2.979)--(11.267,2.979);
\node[gp node right] at (1.320,2.979) {$0.70$};
\draw[gp path] (1.504,3.643)--(1.684,3.643);
\draw[gp path] (11.447,3.643)--(11.267,3.643);
\node[gp node right] at (1.320,3.643) {$0.72$};
\draw[gp path] (1.504,4.308)--(1.684,4.308);
\draw[gp path] (11.447,4.308)--(11.267,4.308);
\node[gp node right] at (1.320,4.308) {$0.74$};
\draw[gp path] (1.504,4.973)--(1.684,4.973);
\draw[gp path] (11.447,4.973)--(11.267,4.973);
\node[gp node right] at (1.320,4.973) {$0.76$};
\draw[gp path] (1.504,5.637)--(1.684,5.637);
\draw[gp path] (11.447,5.637)--(11.267,5.637);
\node[gp node right] at (1.320,5.637) {$0.78$};
\draw[gp path] (1.504,6.302)--(1.684,6.302);
\draw[gp path] (11.447,6.302)--(11.267,6.302);
\node[gp node right] at (1.320,6.302) {$0.80$};
\draw[gp path] (1.504,6.966)--(1.684,6.966);
\draw[gp path] (11.447,6.966)--(11.267,6.966);
\node[gp node right] at (1.320,6.966) {$0.82$};
\draw[gp path] (1.504,7.631)--(1.684,7.631);
\draw[gp path] (11.447,7.631)--(11.267,7.631);
\node[gp node right] at (1.320,7.631) {$0.84$};
\draw[gp path] (1.504,0.985)--(1.504,1.165);
\draw[gp path] (1.504,7.631)--(1.504,7.451);
\node[gp node center] at (1.504,0.677) {$1.3$};
\draw[gp path] (3.161,0.985)--(3.161,1.165);
\draw[gp path] (3.161,7.631)--(3.161,7.451);
\node[gp node center] at (3.161,0.677) {$1.4$};
\draw[gp path] (4.818,0.985)--(4.818,1.165);
\draw[gp path] (4.818,7.631)--(4.818,7.451);
\node[gp node center] at (4.818,0.677) {$1.5$};
\draw[gp path] (6.476,0.985)--(6.476,1.165);
\draw[gp path] (6.476,7.631)--(6.476,7.451);
\node[gp node center] at (6.476,0.677) {$1.6$};
\draw[gp path] (8.133,0.985)--(8.133,1.165);
\draw[gp path] (8.133,7.631)--(8.133,7.451);
\node[gp node center] at (8.133,0.677) {$1.7$};
\draw[gp path] (9.790,0.985)--(9.790,1.165);
\draw[gp path] (9.790,7.631)--(9.790,7.451);
\node[gp node center] at (9.790,0.677) {$1.8$};
\draw[gp path] (11.447,0.985)--(11.447,1.165);
\draw[gp path] (11.447,7.631)--(11.447,7.451);
\node[gp node center] at (11.447,0.677) {$1.9$};
\draw[gp path] (1.504,7.631)--(1.504,0.985)--(11.447,0.985)--(11.447,7.631)--cycle;
\node[gp node center,rotate=-270] at (0.246,4.308) {$Z_\mathrm A$};
\node[gp node center] at (6.475,0.215) {$g_0^2$};
\node[gp node left] at (2.972,1.390) {Padé fit};
\gpcolor{rgb color={0.000,0.000,1.000}}
\gpsetlinetype{gp lt plot 0}
\gpsetlinewidth{2.00}
\draw[gp path] (1.872,1.390)--(2.788,1.390);
\draw[gp path] (1.504,7.531)--(1.591,7.506)--(1.677,7.481)--(1.764,7.455)--(1.851,7.430)%
  --(1.938,7.404)--(2.024,7.379)--(2.111,7.353)--(2.198,7.327)--(2.285,7.301)--(2.371,7.275)%
  --(2.458,7.249)--(2.545,7.223)--(2.632,7.196)--(2.718,7.170)--(2.805,7.143)--(2.892,7.116)%
  --(2.979,7.089)--(3.065,7.062)--(3.152,7.035)--(3.239,7.008)--(3.326,6.980)--(3.412,6.953)%
  --(3.499,6.925)--(3.586,6.897)--(3.672,6.869)--(3.759,6.840)--(3.846,6.812)--(3.933,6.783)%
  --(4.019,6.754)--(4.106,6.725)--(4.193,6.695)--(4.280,6.666)--(4.366,6.636)--(4.453,6.606)%
  --(4.540,6.576)--(4.627,6.545)--(4.713,6.514)--(4.800,6.483)--(4.887,6.451)--(4.974,6.420)%
  --(5.060,6.388)--(5.147,6.355)--(5.234,6.322)--(5.321,6.289)--(5.407,6.256)--(5.494,6.222)%
  --(5.581,6.187)--(5.667,6.153)--(5.754,6.117)--(5.841,6.082)--(5.928,6.046)--(6.014,6.009)%
  --(6.101,5.972)--(6.188,5.934)--(6.275,5.895)--(6.361,5.856)--(6.448,5.816)--(6.535,5.776)%
  --(6.622,5.734)--(6.708,5.692)--(6.795,5.649)--(6.882,5.605)--(6.969,5.561)--(7.055,5.515)%
  --(7.142,5.468)--(7.229,5.420)--(7.315,5.371)--(7.402,5.320)--(7.489,5.268)--(7.576,5.215)%
  --(7.662,5.160)--(7.749,5.103)--(7.836,5.044)--(7.923,4.983)--(8.009,4.921)--(8.096,4.855)%
  --(8.183,4.788)--(8.270,4.717)--(8.356,4.644)--(8.443,4.567)--(8.530,4.486)--(8.617,4.402)%
  --(8.703,4.313)--(8.790,4.220)--(8.877,4.121)--(8.964,4.016)--(9.050,3.904)--(9.137,3.785)%
  --(9.224,3.657)--(9.310,3.519)--(9.397,3.371)--(9.484,3.210)--(9.571,3.034)--(9.657,2.842)%
  --(9.744,2.629)--(9.831,2.393)--(9.918,2.129)--(10.004,1.830)--(10.091,1.489);
\gpcolor{color=gp lt color border}
\node[gp node left] at (2.972,1.840) {$\za$};
\gpcolor{rgb color={1.000,0.000,0.000}}
\draw[gp path] (1.872,1.840)--(2.788,1.840);
\draw[gp path] (1.872,1.930)--(1.872,1.750);
\draw[gp path] (2.788,1.930)--(2.788,1.750);
\draw[gp path] (10.091,1.194)--(10.091,1.792);
\draw[gp path] (10.001,1.194)--(10.181,1.194);
\draw[gp path] (10.001,1.792)--(10.181,1.792);
\draw[gp path] (6.058,5.931)--(6.058,6.035);
\draw[gp path] (5.968,5.931)--(6.148,5.931);
\draw[gp path] (5.968,6.035)--(6.148,6.035);
\draw[gp path] (7.009,5.478)--(7.009,5.697);
\draw[gp path] (6.919,5.478)--(7.099,5.478);
\draw[gp path] (6.919,5.697)--(7.099,5.697);
\draw[gp path] (8.272,4.574)--(8.272,4.816);
\draw[gp path] (8.182,4.574)--(8.362,4.574);
\draw[gp path] (8.182,4.816)--(8.362,4.816);
\gpsetpointsize{5.20}
\gppoint{gp mark 6}{(10.091,1.493)}
\gppoint{gp mark 6}{(6.058,5.983)}
\gppoint{gp mark 6}{(7.009,5.587)}
\gppoint{gp mark 6}{(8.272,4.695)}
\gppoint{gp mark 6}{(2.330,1.840)}
\gpcolor{color=gp lt color border}
\gpsetlinetype{gp lt border}
\gpsetlinewidth{1.50}
\draw[gp path] (1.504,7.631)--(1.504,0.985)--(11.447,0.985)--(11.447,7.631)--cycle;
\gpdefrectangularnode{gp plot 1}{\pgfpoint{1.504cm}{0.985cm}}{\pgfpoint{11.447cm}{7.631cm}}
\end{tikzpicture}
  \caption{Plot of our preliminary estimate of $\za$ versus the bare
    coupling~$\gosq$.  The data points and a Padé fit are shown.}
  \label{f:za}
\end{figure}
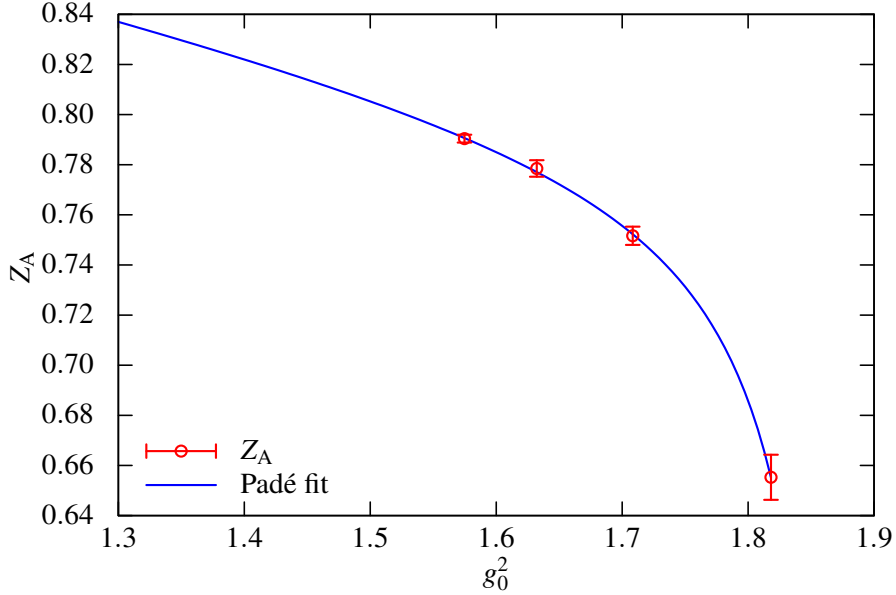

In \fig{f:za}, the $\za$ values of the four ensembles with the
smallest absolute PCAC mass at each value of $\beta$ are plotted
against the bare coupling.  They have been used to determine a Padé
approximation of $\za(\gosq)$ based on the ansatz
\begin{align}
  \label{e:interpolation}
  \za(\gosq) & = \frac{1 + a_1 \cdot \gosq + a_2 \cdot g_0^4}{1 + b_1 \cdot \gosq},
\end{align}
which is constrained to yield the correct continuum limit $\za(0) =
1$.  The coefficients we found are
\begin{align}
  a_1 & = -0.6492, &
  a_2 & =  0.0619, &
  b_1 & = -0.5298.
\end{align}
With these parameters, the fit function lies well within the errors of
all data points.

Let us emphasize again that the results on $\za$ should
be regarded as preliminary, since a final error analysis as well as
the inclusion of the reweighting factors (to compensate for the
approximation errors of the RHMC algorithm employed for the third
quark in our simulations) are still missing.  Regarding the latter,
our experience from $\ca$ leads us to expect only a minor influence
from them.  Moreover, to account for the topology freezing observed at the
finest lattice spacing, we will supplement our definition of $\za$
with the condition to restrict the analysis to the sector of zero
topological charge, as we did for $\ca$~\cite{impr:ca_nf3}.

\section{Conclusions}
\label{s:conclusions}

We have determined a preliminary expression for the renormalization
factor~$\za(\gosq)$ of the isovector axial current for the tree-level
improved gauge action and three dynamical flavors of Wilson fermions.
It is summarized in the interpolation formula of \eq{e:interpolation}.
Together with the improvement coefficient~$\ca$, which has already
been determined non-perturbatively~\cite{impr:ca_nf3}, this will make
it possible to obtain precise results for matrix elements such as
pseudoscalar decay constants.  However, before our result is applied,
we will scrutinize our analysis by projection to the zero-topology
sector and also investigate deviations from the constant physics
condition.  In addition, a simulation at another point ($L/a = 14$,
$\beta = 3.414$) along our line of constant physics is under way to
shed light on the cutoff effects reflected at the coarsest lattice
spacing by the differences between $\za$ and $\zacon$.

\paragraph{Acknowledgments}
We want to thank R.~Hoffmann, P.~Fritzsch and S.~Takeda, who are the
original authors of the script that we used in parts of our analysis.
We are indebted to Rainer Sommer and Stefan Schaefer for helpful
advice as well as Alberto Ramos, who additionally provided us with
his analysis program measuring the gradient-flow coupling.  This work
is supported by the grant HE~4517/3-1 (J.~H.\ and C.~W.) of the
Deutsche Forschungsgemeinschaft.  We gratefully acknowledge the
computing time granted by the John von Neumann Institute for Computing
(NIC) and provided on the supercomputer JUROPA at Jülich
Supercomputing Centre (JSC).  Computer resources were also provided by
DESY, Zeuthen (PAX Cluster), the CERN thqcd2 installation, and the ZIV
of the University of Münster (PALMA HPC cluster).

\bibliographystyle{JHEP_mod}
\bibliography{latticen}

\end{document}